\documentclass[aip,jcp,amsmath,amssymb,preprint,numeric]{revtex4-1}
\usepackage{graphicx}
\usepackage{subfigure}
\usepackage{color}
\usepackage{amsmath}
\usepackage{bm}
\usepackage{multirow}

\begin{document}
\title {A Comparative Study of a Class of Mean Field Theories of the Glass Transition}
\author{Indranil Saha}

\affiliation{\textit{Polymer Science and Engineering Division, CSIR-National Chemical Laboratory, Pune-411008, India}}

\author{Manoj Kumar Nandi}
\affiliation{\textit{Polymer Science and Engineering Division, CSIR-National Chemical Laboratory, Pune-411008, India}}
\author{Chandan Dasgupta}
\affiliation{\textit{Centre for Condensed Matter Theory, Department of Physics, Indian Institute of Science, Bengaluru 560012, India}}
\affiliation{\textit{International Centre for Theoretical Sciences, Bengaluru-560 089, India}}

\author{Sarika Maitra Bhattacharyya}
\email{mb.sarika@ncl.res.in}
\affiliation{\textit{Polymer Science and Engineering Division, CSIR-National Chemical Laboratory, Pune-411008, India}}

\date{\today} 

\begin{abstract}

In a recently developed microscopic mean field theory, we have shown that the dynamics of a system, when described only in terms of its pair structure, can predict the correct dynamical transition temperature. Further, the theory predicted the difference in dynamics of two systems (the Lennard-Jones and the WCA) despite them having quite similar structures. This is in contrast to the Schweizer-Saltzman (SS) formalism which predicted the dynamics of these two systems to be similar. The two theories although similar in spirit have certain differences. Here we present a comparative study of these two formalism to find the origin of the difference in their predictive power. We show that not only the dynamics in the potential energy surface, as described by our earlier study, but also that in the free energy surface, like in the SS theory, can predict the correct dynamical transition temperature. Even an approximate one component version of our theory, similar to the system used in the SS theory, can predict the transition temperature reasonably well. Interestingly, we show here that despite the above mentioned shortcomings the SS theory can actually predict the correct transition temperatures. Thus microscopic mean field theories of this class which express dynamics in terms of the pair structure of the liquid while being unable to predict the actual dynamics of the system are successful in predicting the correct dynamical transition temperature.
\end{abstract}

\maketitle


	
\section{Introduction}

The details of the relaxation dynamics of a glassy system and the properties of a glass continue to be in the focus of intense research activity. These investigations are motivated by the fact that glasses are not only important for many daily and technological applications but are also an intellectual challenge for fundamental studies. One of such challenges is the development of a theoretical framework that can give a satisfactory description of the unusual properties of glassy dynamics. In the normal liquid domain structure plays an important role \cite{hansen2006theory}. Thus there are theories which are developed where the information of the dynamics can be obtained from just the information of the structure of the liquid \cite{gotze2008complex, KAWASAKI2000, Schweizer, wolynes}. However, in the case of supercooled liquids, the dynamics changes over orders of magnitude if the temperature is decreased by a modest amount whereas the structure changes very little. This questions the role of structure in the dynamics. There have been studies showing that two systems, namely the Kob-Andersen model with particles interacting via the Lennard-Jones potential (KALJ) and that via the WCA potential (KAWCA), which have very similar structures, as characterized by the two point correlation functions, have dynamics which are orders of magnitude apart. This result strengthens the idea that the dynamics in the supercooled liquid is dominated by many body correlation \cite{tarjus_prl}. This observation also justifies the findings that theories like the mode coupling theory (MCT) and dynamic density functional theory which requires structure as an input cannot predict the difference in the dynamics of two systems having similar structures\cite{berthier_EPJE}.

However, following these studies, there has been further investigation involving some of us in analyzing the role of pair structure in the dynamics\cite{Atryeeprl, Manojprl}. It was shown that although the structure of the system cannot predict the actual dynamics, it has the information of the dynamical transition temperature, often referred to as the mode coupling transition temperature, $T_{c}$\cite{Atryeeprl, Manojprl}. These studies further showed that the information of the difference in the dynamics of two systems having very similar structure is also embedded in the structure. Small changes in structure can cause a large change in dynamics. Interestingly, the theoretical formulations of these two studies reporting similar observation are quite different. The first one by Banerjee {\it et al.} \cite{Atryeeprl} was based on the phenomenological connection between the relaxation dynamics and the configurational entropy via the well known Adam-Gibbs (AG) relation \cite{adam-gibbs}. In this study, it was shown that the dynamics obtained via AG relation using the pair part of the configurational entropy, which requires only the information of the pair structure diverges at the MCT transition temperature. The study further showed that at the level of this pair dynamics the two systems namely the KALJ and the KAWCA are different. The theoretical framework of the second study by Nandi {\it et al.} \cite{Manojprl} was completely different from the first one\cite{Atryeeprl}. It was a microscopic mean field theory which used the concepts of density functional theory (DFT)\cite{Manojprl}. In this work, starting from exact microscopic many body expression, mean field approximation was made where the mean field incorporated the interaction between the particles at the two body level. The dynamics obtained via the mean first passage time was expressed only as a function of the pair structure of the liquid. The theory when applied for different systems could predict the MCT transition temperatures and also could predict that the dynamics of the KALJ and KAWCA systems are different.

Thus although the two theories are quite different in their approaches, their predictions are quite similar, and in both cases, the dynamics was described by the pair structure of the liquid. Thus these findings redefined the role of pair structure in the dynamics.

Some time back Schweizer and Saltzman have proposed a formalism for obtaining the dynamics in a supercooled liquid \cite{sch_salt_jcp_2003}. Their formalism was also based on DFT and quite similar to the DFT formalism by Nandi {\it et al.} \cite{Manojprl}. However, when Schweizer-Saltzman formalism was applied to KALJ and KAWCA systems, it failed to describe the difference in the dynamics of the two systems \cite{berthier_EPJE}. In order to further develop the formalism proposed by Nandi {\it et al.}, it is important to investigate why two theories which appear to be similar in many ways make different predictions. This is the goal of the present study.

The organization of the paper is as follows. In Section II we describe the systems and the simulation details. This is followed by a comparison of the two theories in Section III. In Section IV we present the MCT fitting procedure used in the present study. The numerical results are presented in Section V followed by conclusions in Section VI.

\section{System and Simulation Details}
This study includes performing extensive molecular dynamics simulations for three-dimensional binary mixtures in the canonical ensemble.  The models studied here are the well-known 
models of glass-forming liquids: the binary Kob-Andersen (KA) Lennard-Jones (LJ) liquids \cite{kob} and the corresponding Weeks-Chandler-Andersen (WCA) version \cite{chandler}. 
The system contains $N_{A}$ particles of type A and $N_{B}$ particles of type B under periodic boundary conditions. The total number density is fixed at $\rho$ = N/V with the total 
number of particles N and a system volume V.

The Kob-Andersen model is a binary mixture of A and B atoms. The system is a 80:20 mixture of A:B (denoted by $\alpha$ and $\beta$, respectively), interacting with a interatomic pair 
potential $U_{\alpha\beta}$(r), described by a shifted and truncated Lennard-Jones (LJ) potential,
\begin{equation} \label{Eq.1}
\begin{split}
U_{\alpha\beta}(r) &=U_{\alpha\beta}^{(LJ)}(r;\sigma_{\alpha\beta},\epsilon_{\alpha\beta})-U_{\alpha\beta}^{(LJ)}(r^{(c)}_{\alpha\beta};\sigma_{\alpha\beta},\epsilon_{\alpha\beta}), r\leq r^{(c)}_{\alpha\beta}\\
&=0, r> r^{(c)}_{\alpha\beta},
\end{split}
\end{equation}
where,
$U_{\alpha\beta}^{(LJ)}(r;\sigma_{\alpha\beta},\epsilon_{\alpha\beta})=4\epsilon_{\alpha\beta}[({\sigma_{\alpha\beta}}/{r})^{12}-({\sigma_{\alpha\beta}}/{r})^{6}]$ and $r^{(c)}_{\alpha\beta}=2.5\sigma_{\alpha\beta}$ for the LJ systems (KALJ) and $r^{(c)}_{\alpha\beta}$ is equal to the position of the minimum of $U_{\alpha\beta}^{(LJ)}$ for the WCA systems(KAWCA). Here length, temperature and time are given in units of $\sigma_{AA}$,
${k_{B}T}/{\epsilon_{AA}}$ and $\surd({m_{A}\sigma_{AA}^2}/{\epsilon_{AA}})$, 
respectively. We have simulated the Kob-Andersen model with the interaction parameters  $\sigma_{AA}$ = 1.0, $\sigma_{BB}$ =0.88 , $\sigma_{AB}=0.8$, $\epsilon_{AA}$ =1.0, $\epsilon_{AB}$ =1.5, $\epsilon_{BB}$ =0.5, $m_{A} =m_{B}$=1.0. \vspace{2mm}

The molecular dynamics (MD) simulations have been carried out using the LAMMPS package \cite{lammps}. We have performed MD simulations in the isothermal canonical ensemble (NVT) using  $Nos\acute{e}$-Hoover thermostat with integration timestep 0.005$\tau$. 
The System size is $N =N_{A}+N_{B}$= 1000 with $N_{A}$ = 800 and $N_{B}$ = 200 (N= total number of particles, $N_{A/B}$= number of particles of type-A/B), here the system has been 
studied at the density $\rho$= 1.2. The system is kept in a cubic box with periodic boundary condition. The time constants for $Nos\acute{e}$-Hoover thermostat is taken to be 100 
timesteps. For all state points, three to five independent samples with run lengths more than \textcolor{blue}{100$\tau_{\alpha}$ ($\tau_{\alpha}$ is the $\alpha$-relaxation time)} are analyzed. \\

The partial structure factors $S_{\mu \nu}(K)$, calculated from simulation studies and used as an input in our theoretical models, can be defined as,
\begin{equation}
S_{\mu \nu}(k)= \frac{1}{\sqrt{N_{\nu}N_{\mu}}} \sum_{i=1}^{N_{\nu}} \sum_{j=1}^{N_{\mu}}
\exp(-i\textbf{k}.(\textbf{r}^\nu_{i}-\textbf{r}^\mu_{j})).
\label{Eq.2}
\end{equation} 

The alpha relaxation time, $\tau_{\alpha}$ has been calculated from the decay of the overlap function $q(t)$, using $q(t = \tau_{\alpha},T)$/N = 1/e. The overlap function is a 
two-point time correlation function of local density $\rho(r,t)$ and it has been used many times in recent studies \cite{shila-jcp}. The function can be represented as,  

\begin{equation}
\begin{aligned}
\langle q(t) \rangle & = \Bigg \langle \sum_{i=1}^N \omega(\mid \textbf{r}_{i}(t_{0})- \textbf{r}_{i}(t+t_{0}) \mid) \Bigg \rangle,
\\ \omega(x) &= 1,x\leq a, \ implying \ overlap \\
&= 0, otherwise.
\end{aligned}
\label{Eq.3}
\end{equation} 

Here $\omega$ defines the condition of overlap between two particle positions separated by a time interval t. The time-dependent overlap function thus depends on the choice of the cutoff parameter a, which we choose to be 0.3.

\section{Theory}

In this section, we compare the recently developed methodology in our group \cite{Manojprl} with that developed earlier by Schweizer and Saltzman \cite{sch_salt_jcp_2003, Schweizer}. 
In our earlier work, we started from an exact microscopic Fokker-Planck expression and made mean field approximations and showed that the dynamics of the system can be described by the mean first passage time of escape from the trapping potential. In the proposed theory of Schweizer-Saltzman, they showed that for dense liquids or suspensions the particle dynamics could also be viewed from the point of a stochastic nonlinear Langevin equation of motion where non-equilibrium free energy governs the physics. Finally, the standard Kramers Theory was used to calculate the dynamics. Both of these approaches used the concept of the density functional theory (DFT). In the approach by Schweizer-Saltzman the free energy was derived using concepts of both idealized mode coupling theory (IMCT) \cite{wolynes} and DFT whereas in our earlier work the mean-field potential was derived using the concepts of DFT. It is also noteworthy that one can always recast any Langevin equation onto a Fokker Planck equation and vice versa. Likewise one can even go from mean first passage time (mfpt) dynamics to Kramers dynamics \cite{zwanzig2001nonequilibrium}. Thus, it makes these two approaches very similar. However it was found that our theory can distinguish between the dynamics of the LJ and WCA systems \cite{Manojprl} whereas the theory developed by Schweizer-Saltzman cannot 
\cite{sch_salt_jcp_2003,berthier_EPJE}. 

In this article, we present a study which is similar in spirit to that presented by Schweizer-Saltzman but for a binary system. However, there are certain differences in the treatment of the problem which will be elaborated in this section. Similar in spirit to Schweizer-Saltzman formalism we start with the over-damped Langevin equation to describe the dynamics of a particle at position `r' in a field `F(r)',  
\begin{equation}
\frac{\partial r}{\partial t} = [-\frac{1}{\zeta}\nabla F(r)+\sqrt{2D}\eta(t)].  
\label{Eq.5} 
\end{equation}
\noindent
Here, $D= \frac{k_{B}T} {\zeta}$; D is the diffusion coefficient, $\zeta$ is the short time friction felt by the particle, $\eta(t)$ is a Gaussian white noise 
satisfying $\langle\eta(t) \rangle$=0 and the fluctuation-dissipation theorem,

\begin{equation}
\langle \eta{(t)} \eta{(t')} \rangle = 2\zeta k_{B}T\delta(t-t').
\label{Eq.6}
\end{equation}

The stochastic Langevin equation in Eq.\ref{Eq.5} with Eq.\ref{Eq.6} can also be written in terms of a probabilistic Fokker-Planck equation as, 

\begin{equation}
\frac{\partial P(r,t)}{\partial t} =
\nabla.[\frac{k_{B}T}{\zeta}\nabla P(r,t)+\frac{P(r,t)}{\zeta}\nabla F(r)].
\label{Eq.7}
\end{equation}
\noindent
Here $P(r,t)$ is the probability distribution function. Upon replacing the probability distribution by single particle time dependent density, $\rho_{s}(r,t)$ we arrive at the following expression which can also be identified with a Smoluchowski equation,
\begin{equation}
\frac{\partial \rho_{s}(r,t)}{\partial t} = \nabla.[\frac{1}{\zeta}(k_{B}T \nabla \rho_{s}(r,t)+\rho_{s}(r,t) \nabla F({r}))].
\label{Eq.8}
\end{equation}  
\noindent
Note that although we have started with a Langevin dynamics we now arrive at the Smoluchowski dynamics of the particle in a field, $F(r)$, which following Schweizer-Saltzman formalism is the effective free energy surface. In our earlier study to describe the dynamics of the system we had arrived at a similar Smoluchowski dynamics but in a potential energy surface\cite{Manojprl}. For a binary system the field is given by $F^{B}(r)$ and can be written as \cite{singh_free},
\begin{equation}
F^{B}(r)=F^{B}_{id}(r)+F^{B}_{ex}(r). 
\label{Eq.9}
\end{equation}
\noindent
Here $F^{B}_{id}(r)$ is the free energy for the binary ideal gas term which can be written as,
\begin{equation}
\beta F^{B}_{id}(r) \simeq -3\ln(r)+(x_{A}lnx_{A}+ x_{B}lnx_B).
\label{Eq.10}
\end{equation}
\noindent
$x_{A/B}$, is the mole fraction of component $A/B$. In the above expression apart from the mixing entropy term following the Schweizer-Saltzman formulation we have retained only the terms which are dependent on `r'. Next we use the density functional theory (DFT) approach of Ramakrishnan-Yussouff (RY) \cite{RY_theory} to get the excess part of the free energy $F^{B}_{ex}$, 
\begin{equation}
\begin{aligned}
\beta F_{ex}^{B}(\rho{(\textbf{R,t})}) \simeq& -\frac{1}{2} \int d\textbf{R} \int d\textbf{R}' \sum_{\alpha \beta} \rho_{\alpha}(\textbf{R,t})C_{\alpha \beta}(|\textbf{R}-\textbf{R}'|)\rho_{\beta}(\textbf{R}',t) \\
&= -\frac{1}{2} \int \frac{d\textbf{q}}{(2\pi)^3} \sum_{\mu \nu} C_{\mu \nu}(q) \rho_{\mu}(\textbf{-q},t) \rho_{\nu}(\textbf{q},t), 
\label{Eq.11}
\end{aligned}
\end{equation}    
\noindent
The density $\rho(\textbf{R},t)$ = $ \sum_{i} \delta(\textbf{R}-\textbf{R}_{i}(t))$ and $C(|\textbf{R}-\textbf{R}'|)$ is the direct correlation function, where \textbf{R} denotes the position of the particle. $\rho(\textbf{q},t)$ and $C(\textbf{q},t)$ are the density and direct correlation function in the wave number space respectively. Next we describe how we use this expression of free energy in Eq.\ref{Eq.9} to solve Eq.\ref{Eq.8}. In Eq.\ref{Eq.11} the excess free energy is dependent on the density of the particle. When this expression of free energy is put back in Eq.\ref{Eq.8} then it requires an iterative solution of the time dependent density \cite{Archer_pre, Hopkins_jcp}. However, in this work we do not use the iterative method but make certain standard approximations \cite{wolynes,Schweizer,Manojprl} which allow us to express the free energy as a function of the structure of a liquid and a displacement parameter. The steps followed here are, first we make the mean field like Vineyard approximation. Thus, we assume that $\rho(R,t) = \rho(R(0)) \rho_s(R,t)$. The Vineyard approximation is reasonable as we are dealing with short term cage breaking dynamics where both the single particle and the collective dynamics are similar. To dynamically close the theory at the single particle level we write, $ \rho_s(R,t)$ $\simeq$ $(\alpha_{i}(t) /\pi)^{3/2} \exp(-R^2 \alpha_{i}(t))$. At the level of $\alpha(t)$ we do not differentiate between the two species thus we write $\alpha_{i}(t)=\alpha(t)$. We obtain $\rho_s{(q,t)} = \exp(-q^2/4\alpha(t))$. Note that we ignore the self term (i=j) in Eq.10 as we are only interested in the effective interaction from all the other (N-1) particles. With these approximations the excess part of the free energy can be written as,
	\begin{equation}
	\beta F_{ex}^B(\alpha(t)) \simeq -\frac{1}{2} \int
	{d\textbf{q}} \sum_{\mu\nu}C_{\mu \nu}{(q)} [N^{-1} \sum_{(i \neq j)}^{N_{\mu}} \sum^{N_{\nu}} e^{-i\textbf{q}.(\textbf{R}^{\mu}_{i}(0)-\textbf{R}^{\nu}_{j}(0))}] e^{-[\frac{q^2}{2 \alpha(t)}]}.
	\label{Eq.12}
	\end{equation}
	\noindent
	Here, $N_{\mu}$ and $N_{\nu}$ are the number of the ${\mu}$ and ${\nu}$ type of particles respectively and as mentioned earlier we assume the `$\alpha$' to be same for both of them.
	Next, we write, $[N^{-1} \sum_{(i \neq j)}^{N_{\mu}} \sum^{N_{\nu}} e^{-i\textbf{q}.(\textbf{R}^{\mu}_{i}(0)-\textbf{R}^{\nu}_{j}(0))}]$=$\rho x_{\mu} x_{\nu} h_{\mu \nu}(q)$, where $x_{\mu}$/$x_{\nu}$ is the mole fraction of component $\mu$/$\nu$ in the mixture. Thus $F^{B}_{ex}$ can be written as, 
	\begin{equation}
	\beta F_{ex}^B(\alpha(t)) \simeq -\frac{1}{2} \int d\textbf{q} \sum_{\mu \nu} C_{\mu \nu}{(q)} \rho x_{\mu} x_{\nu} h_{\mu \nu}(q) e^{-[\frac{q^2}{2 \alpha(t)}]}.
	\label{Eq.13}
	\end{equation}
	\noindent
	In the above equation the term, $h_{\mu \nu}(q)$ is calculated from simulated partial static structures $\rho x_{\mu} x_{\nu} h_{\mu \nu}(q)$ =$(S_{\mu \nu}(q)-\delta_{\mu \nu})$ and $ \textbf{C}{(q)} $=$(\textbf{1}-{\textbf{S}^{-1}(q)})$, where \textbf{C}{(q)}, \textbf{1}, \textbf{S}(q) are all written as matrix. Note that $ \alpha(t) = \frac{3}{2\langle r^{2}(t)\rangle}$, where $\langle r^{2}(t)\rangle$ is the mean-square displacement. In this formalism, the correlation between the particles exist only at t=0 (given in the vertex of eq.12) and it describes the field in which the particle moves. Once that is taken into account, we can consider each particle to be independent and moving in the above mentioned field. Thus we can now consider the initial position of a particle to be at the origin and in this body fixed frame the displacement of the particle becomes its position. So, one can assume $\langle r^{2}(t)\rangle = r^{2}(t)$. Thus, the expression for free energy can be written as, 
	\begin{equation}
	\beta F_{ex}^B(r(t)) \simeq -\frac{1}{2} \int d\textbf{q} \sum_{\mu \nu} C_{\mu \nu}{(q)} \rho x_{\mu} x_{\nu} h_{\mu \nu}(q) e^{-[\frac{q^2 r^2(t)}{3}]}.
	\label{Eq.28}
	\end{equation}

Now, the overall form of $F^{B}({r(t)})$ is,

\begin{equation}
\beta F^B(r(t)) \simeq -3\ln(r(t))+(x_{\mu}ln(x_{\mu})+ x_{\nu}ln(x_{\nu}))-\frac{1}{2} \int d\textbf{q} \sum_{\mu \nu} C_{\mu \nu}{(q)} \rho x_{\mu} x_{\nu} h_{\mu \nu}(q) e^{-[\frac{q^2 r^2(t)}{3}]}.
\label{Eq.14}
\end{equation}
\noindent

The free energy has two competing terms. The ideal part favours delocalization and fluidization of the system whereas the excess part which arises due to interaction between particles, as is evident from the presence of the direct correlation function in the third term in the r.h.s., traps the particle and thus favours localization. These two competing terms at certain temperature gives rise to a minimum when plotted against `$r$' and this minimum becomes deeper as the temperature is lowered.  Since our system is isotropic the $F^{B}(r(t))$ is dependent only on the scalar value of `r'  so Eq.\ref{Eq.8} reduces to an one-dimensional form which now can be written as, 
\begin{equation}
\frac{\partial \rho(r,t)}{\partial t} = \frac{k_{B}T}{\zeta} \frac{\partial}{\partial{r}}e^{-\beta F^{B}(r(t))} \frac{\partial}{\partial{r}}e^{\beta F^{B}(r(t))} ,
\label{Eq.15}
\end{equation}  
\noindent
which can also be written as, 
\begin{equation}
\frac{\partial \rho(r,t)}{\partial t} = \mathcal{L} \rho(r,t).
\label{Eq.16}
\end{equation}
\noindent
From the above Smoluchowski equation where, $ 
\mathcal{L} = \frac{k_{B}T}{\zeta} \frac{\partial}{\partial{r}}e^{-\beta F^{B}(r(t))} \frac{\partial}{\partial{r}}e^{\beta F^{B}(r(t))}$ using standard formalism \cite{zwanzig}, we can calculate the mean first passage time for the binary system $\tau^B_{mfpt}$, which is the time taken by particles to leave the caging created by the 'nonequilibrium free energy' i.e. $F^{B}(r(t))$,
\begin{equation}
\begin{split}
\mathcal{L}^\dagger \tau^B_{mfpt} = -1  \\ \\
D_{0} e^{\beta F^{B}(r(t))} \frac{\partial}{\partial{r}}e^{-\beta F^{B}(r(t))} \frac{\partial}{\partial{r}}{\tau^B_{mfpt}}  =-1 \\ \\ 
\tau^B_{mfpt} = \frac{1}{D_{0}} \int_{0}^{r_{max}} e^{\beta F^{B}(y)} dy \int_{0}^{y} e^{-\beta F^{B}(z)} dz .
\end{split}
\label{Eq.17}
\end{equation}
\vspace{2mm}
\noindent
Here $\mathcal{L}^\dagger$ is adjoint Smoluchowski operator, $D_{0}$ is the diffusion coefficient, $r_{max}$ is the position of the maximum in $F^{B}(r(t))$. At this point let us compare the derivation of the present mean first passage time to that derived in an earlier work \cite{Manojprl}. As mentioned before, in the earlier work we started from the exact microscopic expression which has the full many body correlation and then the mean field approximation was made to calculate the approximate dynamics. The mean field potential which included the information of the interaction between the particles was described by the pair structure of the liquid \cite{correction}. The present formalism starts with writing the dynamics of a particle in an effective free energy surface. Thus unlike in the earlier work here we do not actively require to make any mean field approximation. However, the free energy is obtained using a similar mean field approximation and includes the information of the interaction between particles and is expressed in terms of the pair structure of the liquid. Thus the earlier work and the present work are not identical but similar in spirit.

Next, we show, as discussed by Zwanzig \cite{zwanzig}, under certain approximations the mean first passage time in Eq.\ref{Eq.17} can be written in terms of Kramers first passage time \cite{KRAMERS1940284}. In the above derivation when $k_{B}T$ is small, the integral over z is dominated by the minimum on the free energy surface. We expand $F^{B}(z)$ in quadratic form, $F^{B}(z)= F^{B}_{min}+\frac{1}{2} \omega^2_{min}(z-z_{min})^2$+.... and the upper limit of the first integration is replaced by infinity. Also the integral over y is dominated near the barrier and similar quadratic expansion can be done for $ F^{B}(y)= F^{B}_{max}-\frac{1}{2} \omega^2_{max}(y-y_{max})^2$+.... Note that since we are in the spherical polar co-ordinate the minimum value of y is zero. Also since we assume that the particle is at the minimum, we can replace $x_{max}$ by infinity. The integration thus yield the following expression for $\tau^{B}_{mfpt}$,

\begin{equation}
\begin{aligned}
\tau_{mfpt}^{B}  &= \frac{1}{D_0}\Bigg (\frac{1}{2} \sqrt{\frac{2\pi k_{B}T}{\omega^2_{max}}}  e^{\beta {F^{B}_{max}}} \Bigg) \Bigg ( \frac{1}{2} \sqrt{\frac{2\pi k_{B}T}{\omega^2_{min}}}
e^{-{\beta {F^{B}_{min}}}} \Bigg) \\
&= (\frac{\pi k_{B}T}{2{D_{0}}\omega_{max}\omega_{min}})  e^{ \Big (\beta {F^{B}_{max}-\beta F^{B}_{min}} \Big)} \\ 
&= \tau_{0} e^{\beta [{\Delta F^{B}(T)}]} 
= \tau_{Kramers}.
\end{aligned}
\label{Eq.18}
\end{equation}
\vspace{2mm}
\noindent
Here ${\Delta F^{B}(T)}$=${F^{B}_{max}-F^{B}_{min}}$ i.e. the height of the barrier. Thus the mean first passage time obtained from Smoluchowski equation in certain limit is similar to the Kramers first passage time. 

For the sake of clarity in the next part we first recapitulate the Schweizer-Saltzman formulation \cite{Schweizer,sch_2006, sch_salt_jcp_2003} and then work on it further. Here, we mostly retain the form of the equations and name of the variables used in the work of Schweizer-Saltzman and towards the end mention the similarity between these variables and the ones used in our present study. The main idea of the Schweizer-Saltzman theory is to describe the dynamics in the non-equilibrium free energy surface ($F_{eff}(r)$).
Thus, for an overdamped case the non linear Langevin equation can be written as,

\begin{equation}
-\zeta \frac{\partial r}{ \partial t} -\frac{\partial{F_{eff}}}{\partial r} + \delta{f}= 0.
\label{Eq.19}
\end{equation}
\noindent
The random force also obeys the following white-noise relation,

\begin{equation}
\langle\delta f(t) \delta f(t')\rangle=2\zeta k_{B}T\delta(t-t').
\label{Eq.20}
\end{equation} 
\noindent

Note these expressions are similar to our starting expressions (Eq.\ref{Eq.5} and Eq.\ref{Eq.6}). The free energy $(F_{eff}(r))$ can be written as sum of two parts, the `ideal' term $(F_{0}(r))$ and the `interaction' term $(F_{I}(r))$. It is represented as \cite{singh_free},

\begin{equation}
F_{eff}(r) \equiv F_{0}(r)+F_{I}(r).
\label{Eq.21}
\end{equation}
\noindent
The interaction term $F_{I}$ was heuristically derived to ensure that the force obtained from the free energy matches the idealized mode-coupling theory (IMCT) expression and it was written as, \cite{sch_salt_jcp_2003}
\begin{equation}
\beta F_{eff}(r) \equiv -3\ln(r)-\int \frac{dq}{(2\pi)^3}\rho C^2{(q)}S(q)[1+S^{-1}(q)]^{-1}\exp-[\frac{q^2r^2}{6}(1+S^{-1}(q))].
\label{Eq.22}
\end{equation}
\noindent
In the IMCT expression both self and collective structure factors are present which give rise to the $[1+S^{-1}(q)]$ term in the exponential. However the $[1+S^{-1}(q)]$ term in the vertex of the free energy is present to make sure that its derivative is similar to the IMCT expression. No such term can be present in the vertex in our free energy expression even if we avoid the Vineyard approximation. However if we make the Vineyard approximation in IMCT then we can write $[1+S^{-1}(q)]\simeq2$ both in the vertex and in the exponential of the interaction part of the free energy. Thus $F_{I}(r) \simeq F^{One}_{ex}(r)$,
where $F^{One}_{ex}(r)$ is,

\begin{equation}
\beta F^{One}(r)= \beta F_{0}(r)+\beta F_{ex}(r) \equiv -3ln(r)-\frac{1}{2}\int \frac{dq}{(2\pi)^3}\rho C^2{(q)}S(q)\exp-[\frac{q^2r^2}{3}].
\label{Eq.23}
\end{equation}        
Note that this above expression is the one component version of Eq.\ref{Eq.28}.
\\
Next Schweizer-Saltzman\cite{Schweizer} used the Kramers theory of the mean-first passage time to obtain the dynamics of the Langevin equation (Eq.\ref{Eq.19}). The relaxation time ($\tau^{SS}_{Kramers}$) here is the hopping time that is needed for the particle to escape from localization in the minimum,

\begin{equation}
\tau^{SS}_{Kramers}(T) = \tau_{0} e^{\beta [{\Delta{F_{eff}(T)}}]}.
\label{Eq.24}
\end{equation}
\noindent
Here ${\Delta{F_{eff}(T)}}$ is the barrier height of the free  energy surface created due to the two competing terms in Eq.\ref{Eq.22}. Similar to Eq.\ref{Eq.14} at a smaller value of `r' it will have localization, and at a higher value of `r', it will behave like an ideal gas. Also, in Eq.\ref{Eq.24} $\tau_{0}$ is a prefactor taken as constant as it is weakly dependent on the temperature and density. It also includes the information about the curvature around the maximum and minimum. 
Now, similar to earlier calculation (Eq.\ref{Eq.5}-Eq.\ref{Eq.8}) we can write an equivalent Fokker-Planck equation for Eq.\ref{Eq.19} and calculate the $\tau_{mfpt}$. The expression thus obtained is,
\begin{equation}
\tau^{SS}_{mfpt}= \frac{1}{D_0} \int_{0}^{r_{max}} e^{\beta F_{eff}(y)} dy \int_{0}^{y} e^{-\beta F_{eff}(z)} dz,
\label{Eq.25}
\end{equation}
\noindent
where $D_0$ = $\frac{k_{B}T}{\zeta}$ and $r_{max}$ is the position of the maxima. Interestingly for Eq.\ref{Eq.25} if we follow the steps used for Eq.\ref{Eq.17} to arrive at Eq.\ref{Eq.18} we will get back Eq.\ref{Eq.24}. We can also obtain the dynamics from the one component version of our binary system, 
\begin{equation}
\tau^{One}_{mfpt}= \frac{1}{D_0} \int_{0}^{r_{max}} e^{\beta F^{One}(y)} dy \int_{0}^{y} e^{-\beta F^{One}(z)} dz.
\label{Eq.26}
\end{equation}
\noindent

As mentioned before although these two approaches, ours (present and old)  and Schweizer-Saltzman's are similar, but they make different predictions. In our earlier work where we studied the dynamics in the potential energy surface we could differentiate between the LJ and the WCA systems \cite{Manojprl}, however, it was shown that the Schweizer-Saltzman formalism could not differentiate the two systems \cite{berthier_EPJE}.
In the rest of the article, we will analyze the numerical results obtained from these formalism. We will first use the simulated structure factor to enumerate the values of the relaxation times. We will then study if our present formalism can make predictions similar to our earlier work\cite{Manojprl}. We will also analyze what gives rise to the difference in the prediction made by our earlier theory and Schweizer-Saltzman theory.

\section{MCT Power Law Fitting}

In this section, we will discuss the MCT fitting procedure. The techniques and assumptions in fitting the data are somewhat ambiguous. Establishing a uniform fitting method is thus important for our analysis. \\

In the low temperature regime the dynamics follows a power law behaviour, and it can be described by an algebraic divergence \cite{tarjus_pre,szamel_MCTfitting,gotze2008complex},
\begin{equation}     \label{Eq.27}
\tau(T) \sim (T-T_{c})^{-\gamma} = a(\frac{T}{T_{c}}-1)^{-\gamma}.
\end{equation}   
\noindent
where, $\gamma$ is MCT power law exponent, $T_{c}$ is the mode-coupling transition temperature and `a' is the proportionality constant. From Eq.\ref{Eq.27} we can write,
$\ln(\tau)= \ln(a)+(-\gamma)\ln(\frac{T}{T_{c}}-1)$.
We perform a three parameter (a, $\gamma$, $T_{c}$) fit to the relaxation time for different temperature ranges.  \\

Usually it is known \cite{szamel_MCTfitting} that the MCT power-law region is valid below the onset temperature, $T_{onset}$ and above $T_{c}$. However the $T_{onset}$ doesn't have a well-defined \cite{srikanth,Scio-kob-tarta_jpcond,Reichman}and is method dependent and $T_{c}$ is the parameter obtained from the fitting procedure \cite{kob_tc}. Thus, it's difficult to a priori detect the ideal range. It is also known that by varying the temperature range, one can get multiple power-law fits and thus a range of $T_{c}$ and $\gamma$ values. Hence we vary the range of temperatures and to quantitatively evaluate the goodness of fit we calculate a parameter \cite{rice2006mathematical} $\chi^2= \sum_{i}^{N} \frac{(O_{i}-E{i})^2}{E{i}}$, where $E_{i}$ is the expected value for a data point, obtained from the plot of simulated data, $O_{i}$ is the observed fitted value for that same data point taken from the fitted line and N is the number of data points used in the fit. So, the goodness of a fit can be judged from the $\chi^2$ value, and a lower value of $\chi^2$ will correspond to a better fit. Note that since we do not expect MCT fitting in the high temperature region, for the calculation of $\chi^2$ we start from the data which first shows MCT power law behaviour. \\

\begin{figure}[ht!]
	\centering
	\subfigure{
		\includegraphics[width=0.45\textwidth]{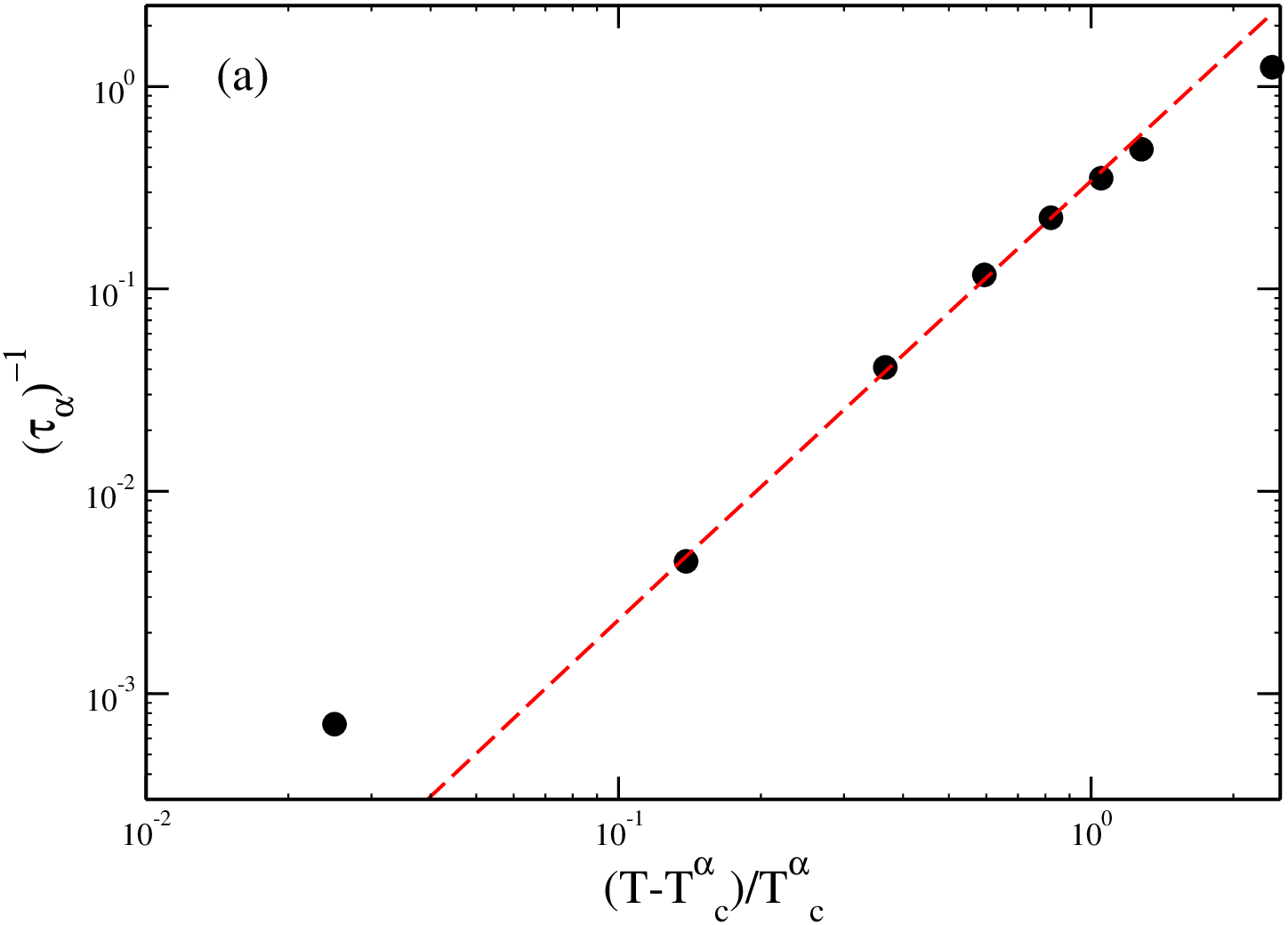}}
	\subfigure{
		\includegraphics[width=0.45\textwidth]{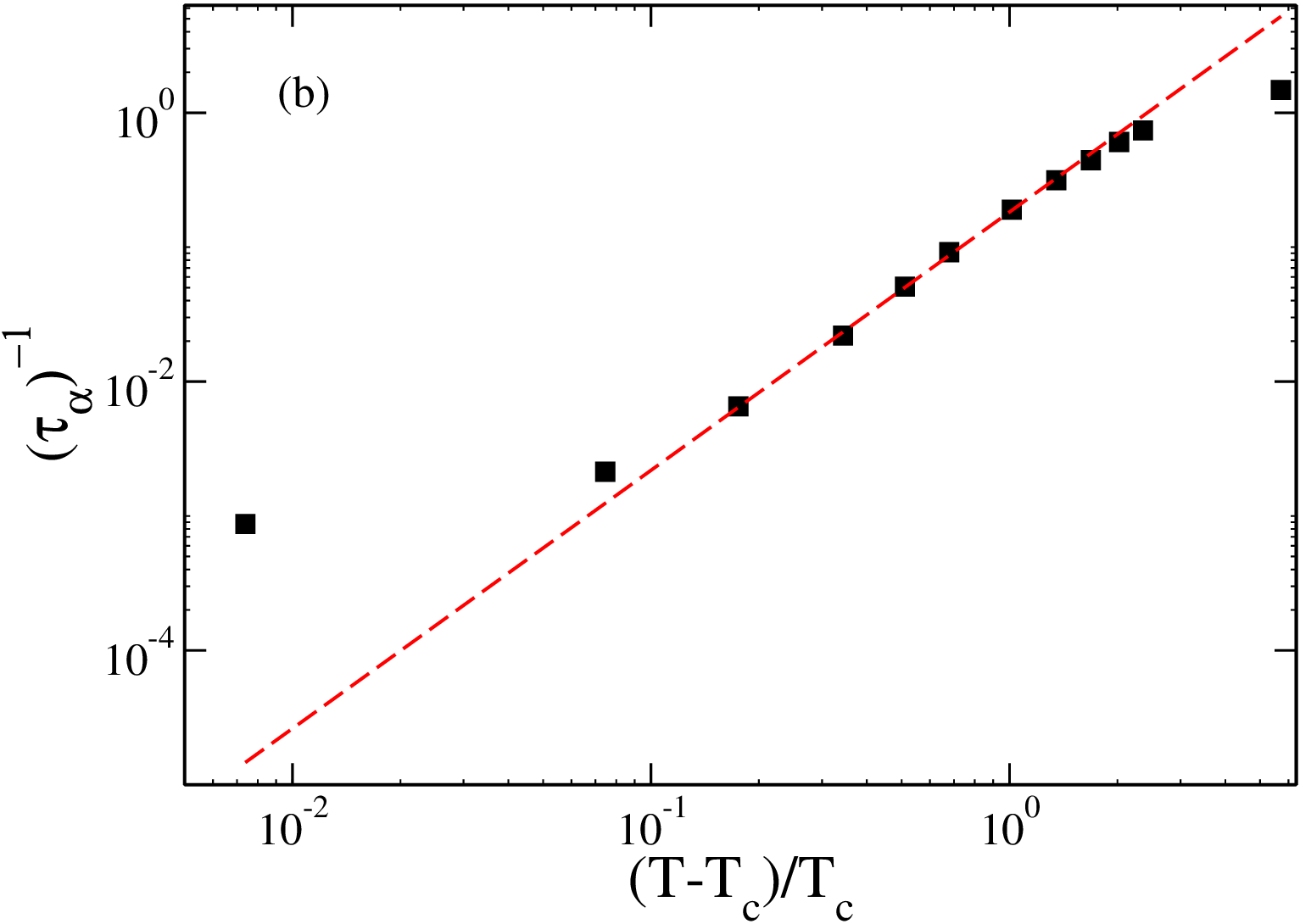}}
	\caption{\it{  MCT power law fit for simulated $\tau_{\alpha}$ values: (a) KALJ and (b) KAWCA systems. In Tables \ref{t1} and \ref{t2} we provide the values of the transition temperatures $T_{c}$ and power law exponents $\gamma$ obtained for different fitting ranges. The plots are for the best fitting regime. For the KALJ system the best fitting is obtained between the temperature range $0.50 \leq T \leq0.90$ where $T_{c}$=0.42 and $\gamma$=2.67. For the KAWCA system the temperature range is $0.35\leq T \leq0.70$ having $T_{c}$=0.29 and $\gamma$=1.92. Here, dashed lines are MCT fits.}}
	\label{fig.1}
\end{figure}

To validate the procedure, we first use it to obtain the fitting parameters for the simulated alpha relaxation time, $\tau_{\alpha}$ of the well known KALJ and KAWCA models. In this work, we obtain the $\tau_{\alpha}$ from the overlap function as the time scale obtained from the decay of the overlap function is known to be the same as or proportional to that obtained from the self-intermediate scattering function.
The results are given in Table \ref{t1} and Table \ref{t2} which shows the $T_{c}$ and the $\gamma$ values. Fig.\ref{fig.1} has the plot for the fitting parameter having least possible value of $\chi^2$. We note that for the LJ system the best fitting is obtained between the temperature range $0.50 \leq T \leq0.90$ where $T_{c}$=0.42 and $\gamma$=2.67. For the WCA system the temperature range is $0.35\leq T \leq0.70$ predicting $T_{c}$=0.29 and $\gamma$=1.92. The plots clearly show that the MCT region is, $ 10^{-1}< (\frac{(T-T_{c})}{T_{c}})<10^{0}$, \cite{manoj_unravel} which is also another criterion (apart from $\chi^{2}$ value) which will be used as a guiding principle to obtain the  $T_{c}$ and $\gamma$ values. These values of the parameters and the fitting range are similar to that obtained in other studies \cite{szamel_MCTfitting} which validates our present method. \\
\vspace{5mm}
\noindent

\begin{table}
	\begin{center}
		\caption{\it{The fitted values of transition temperatures, $T_{c}$, power law exponent, $\gamma$ and quality of the fit, $\chi^2$ (see text) for the simulated $\tau_{\alpha}$ values in KALJ system. The fitting is performed over different temperature regimes for $\rho$ =1.2.  Here the smallest value of $\chi^2$ describes the best fitting.}} 
		\begin{tabular}{ |p{2.5cm}||p{2cm}|p{2cm}|p{2cm}| }
			\hline
			Region(No.of points) & $\chi^2$ & $T_{c}{(LJ)}$ & $\gamma{(LJ)}$ \\
			\hline
			0.45-0.9(6) & $5.26\times10^{-2}$ & 0.4103 & 2.5385 \\
			0.45-0.8(5) & $1.80\times10^{-1}$ & 0.4036 & 2.7443 \\
			0.50-0.9(5) & $1.93\times10^{-2}$ & 0.4239 & 2.6715 \\
			0.50-0.8(4) & $4.01\times10^{-2}$ & 0.4399 & 2.1580 \\
			\hline
		\end{tabular}
		\label{t1}
	\end{center}
\end{table}
\newpage

\begin{table}
	\begin{center}
		\caption{\it{The fitted values of transition temperatures, $T_{c}$, power law exponent, $\gamma$ and quality of the fit, $\chi^2$ (see text) for the simulated $\tau_{\alpha}$ values in KAWCA system. The fitting is performed over different temperature regimes for $\rho$ =1.2.  Here the smallest value of $\chi^2$ describes the best fitting.}}
		\begin{tabular}{ |p{2.5cm}||p{2cm}|p{2cm}|p{2cm}|}
			\hline
			Region(No.of points) & $\chi^2$ & $T_{c}{(WCA)}$ & $\gamma{(WCA)}$ \\
			\hline
			0.32-0.7(7) & $6.87\times10^{-2}$ & 0.2815 & 2.1362 \\
			0.32-0.6(6) & $2.11\times10^{-1}$ & 0.2762 & 2.2837 \\
			0.35-0.7(6) & $2.47\times10^{-2}$ & 0.2978 & 1.9215 \\
			0.35-0.6(5) & $8.02\times10^{-2}$ & 0.2918 & 2.0453 \\
			\hline
		\end{tabular}
		\label{t2}
	\end{center} 
\end{table}
\noindent


\section{Results}

In this section, we present the numerical results for the different first passage times derived in Section III. We also perform MCT fitting for them. First, we discuss the results for the binary systems. Note that in our earlier work we have studied the dynamics on the potential energy surface and have shown that although the dynamics is not the true dynamics of the system, it can predict the known MCT transition temperature value \cite{Manojprl,correction}. We have also shown that the formalism can predict that the dynamics and also the MCT temperatures for the WCA and the LJ systems are different. In this article since we want to compare our formalism with the Schweizer-Saltzman work, we study the dynamics in the free energy surface. We find that similar to our earlier study where the potential energy surface was described by the pair correlation function here the free energy surface is completely described by the pair correlation function (Eq.\ref{Eq.14}). Thus both the present and the earlier studies do not predict the actual dynamics of the system but predict the one described only by the pair correlation function. Also note that, the $\tau_{\alpha}$ measures the time scale of escape from the ``cage" formed by the neighbours. So, it is appropriate to compare this time scale with the mean first passage time for escape from the effective mean-field potential well defined in our calculation. \\

\begin{figure}[ht!]
	\centering
	\subfigure{
		\includegraphics[width=0.45\textwidth]{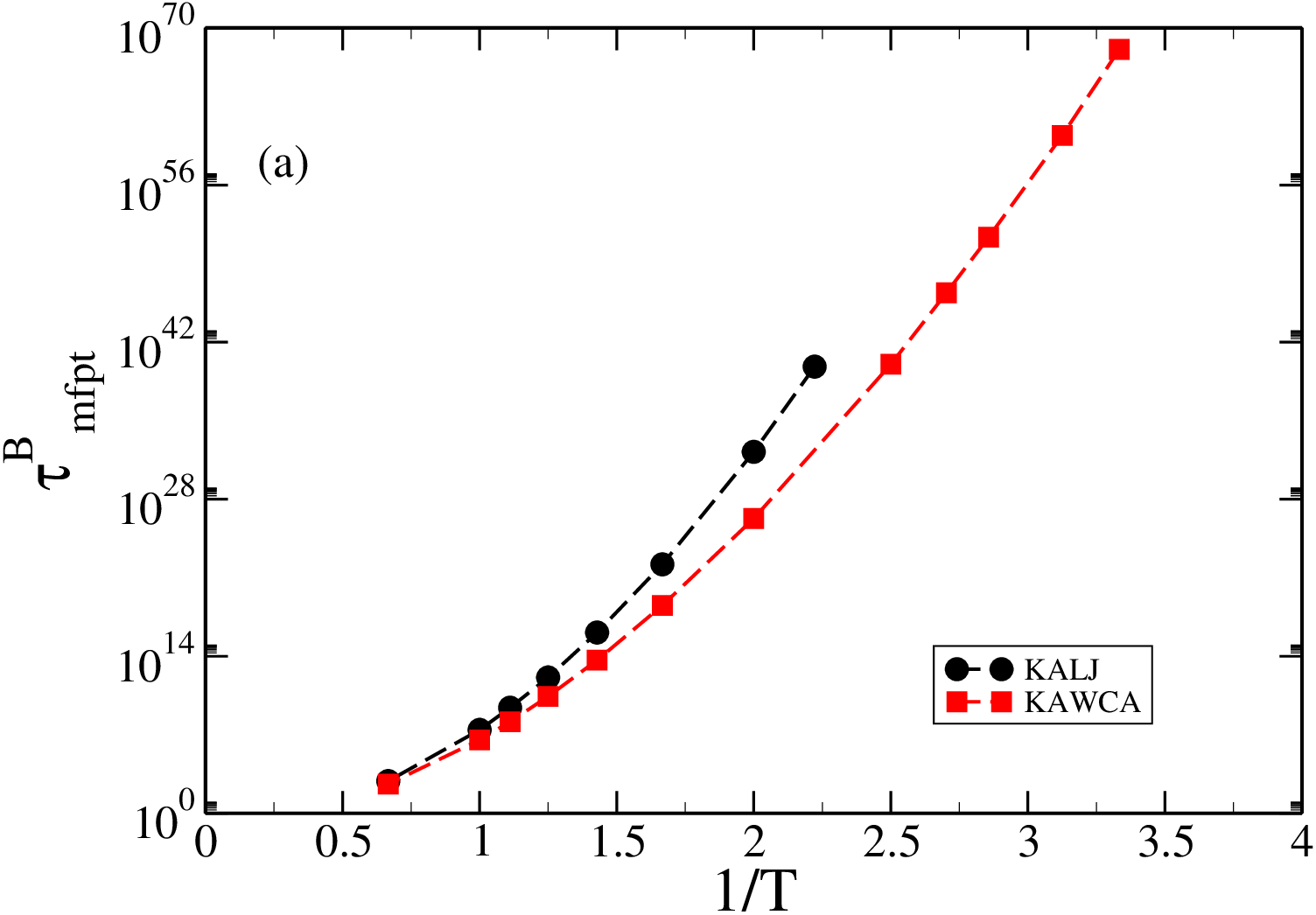}}
	\subfigure{
		\includegraphics[width=0.45\textwidth]{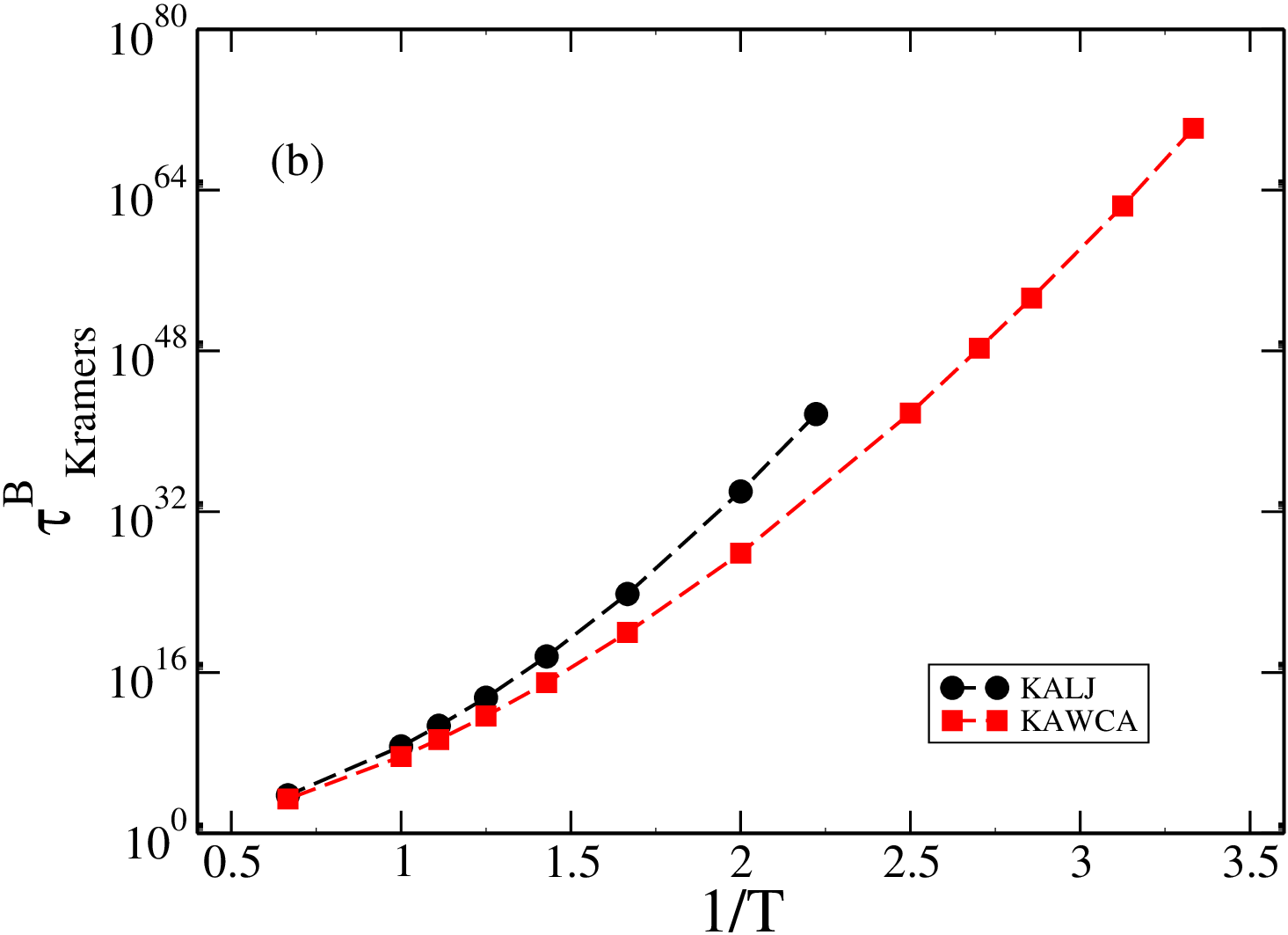}}
	\caption{\it{ Plots for the binary system (a) The Arrhenius plot of $\tau^B_{mfpt}$ (Eq.\ref{Eq.17}) against inverse temperature for KALJ and KAWCA systems at density $\rho=1.2$ for various temperatures. (b) The Arrhenius plot of $\tau^B_{Kramers}$ (Eq.\ref{Eq.18}) against inverse temperature for KALJ and KAWCA systems at density $\rho=1.2$ for various temperatures. }}
	\label{fig.2}
\end{figure}

\begin{figure}[ht!]
	\centering
	\subfigure{
		\includegraphics[width=0.45\textwidth]{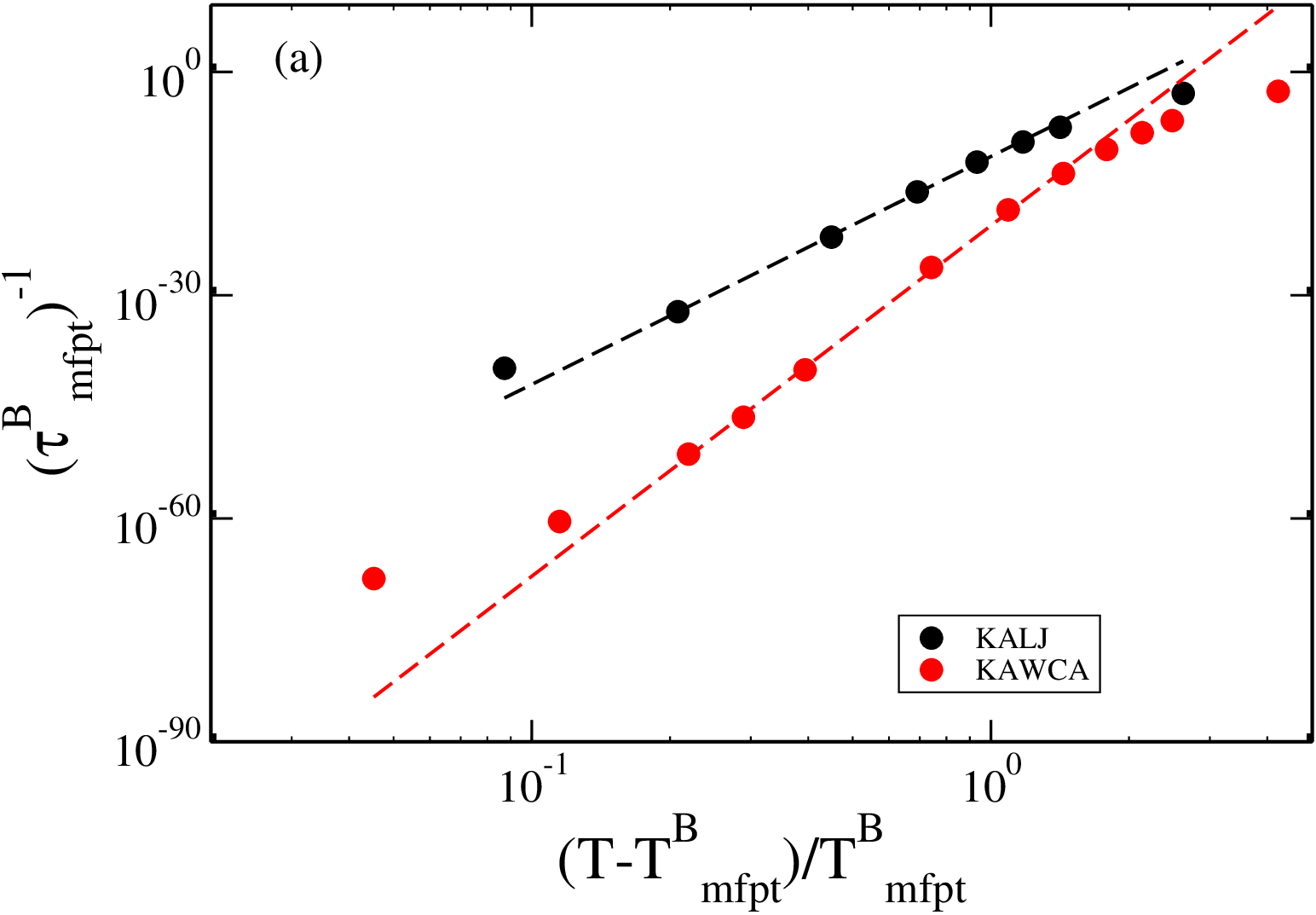}}
	\subfigure{
		\includegraphics[width=0.45\textwidth]{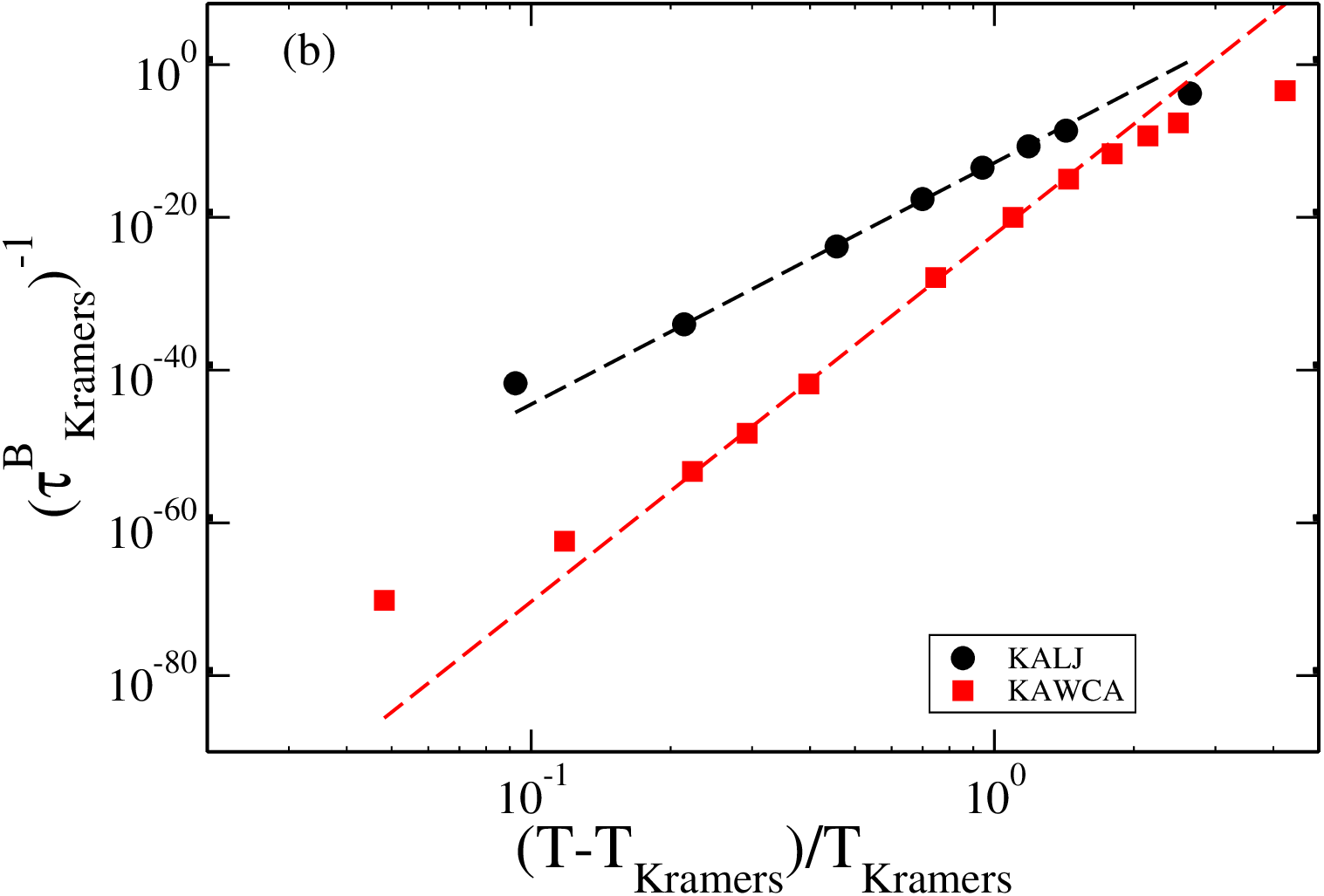}}
	\caption{\it{ Plots for the binary system  (a)  The power law dependence of  $1/{\tau^B_{mfpt}}$ predicts a transition temperature, ${T^B_{mfpt}}$= 0.4139 (LJ) and 0.2870 (WCA). Here, dashed lines are MCT fits. (b) The power law dependence of  $1/{\tau^B_{Kramers}}$ predicts a transition temperature, $T^B_{Kramers}$= 0.4119 (LJ) and 0.2862 (WCA). Here, the dashed lines are MCT fits.}}
	\label{fig.3}
\end{figure}

First, we analyze if the prediction from our present study is similar to that obtained in our earlier study \cite{Manojprl}. In Fig.\ref{fig.2}(a) and Fig.\ref{fig.2}(b) we plot the values of the relaxation times for the binary KALJ and KAWCA systems, as obtained from the mean first passage time and from the Kramers first passage time, respectively. Note that the values of $\tau^{B}_{mfpt}$ and $\tau^{B}_{Kramers}$ are not identical because in $\tau^{B}_{Kramers}$ we take the value of $\tau_{o}$ to be unity however in $\tau^{B}_{mfpt}$ the contribution of $\tau_{o}$ is incorporated via the integration. Although the values of the two first passage times are different, their natures are quite similar. In both the cases, we show that we can differentiate between the dynamics of the  KALJ and the KAWCA systems. Next, we perform the MCT fitting of $\tau^{B}_{mfpt}$ and $\tau^{B}_{Kramers}$ for both the systems. In Tables \ref{t3}-\ref{t6} we first present the different fitting zones used in our analysis and also the respective $\chi^{2}$ values and the corresponding $T_{c}$ and $\gamma$ values. In Fig.\ref{fig.3} we show the fittings which have the minimum $\chi^{2}$ values. We find that both the first passage times can predict transition temperatures which are close to the MCT $T_{c}$ values of the two systems. However, since this dynamics is not the actual dynamics of the system but that predicted via only the pair correlation function thus it cannot predict the correct $\gamma$ values. Thus the present study of the dynamics in the free energy surface and the earlier study of the dynamics in the potential energy surface \cite{Manojprl} provide similar results. This observation is also quite similar to that obtained in an earlier study by some of us \cite{Atryeeprl}. We found that relaxation dynamics obtained via the well known Adam-Gibbs relation, using the information of configurational entropy only at the two body level can predict the correct MCT transition temperature. However, the dynamics thus obtained has a stronger temperature dependence as compared to the actual dynamics of the system. It is noteworthy that in our formalism the cage is static; however, in reality, the cage is dynamic. The time-scale of escape from a static cage is much higher than that of a dynamic cage. Thus, the mfpt dynamics is slower than the $\alpha$ relaxation process. \\

\begin{figure}[ht!]
	\centering
	\includegraphics[width=0.70\textwidth]{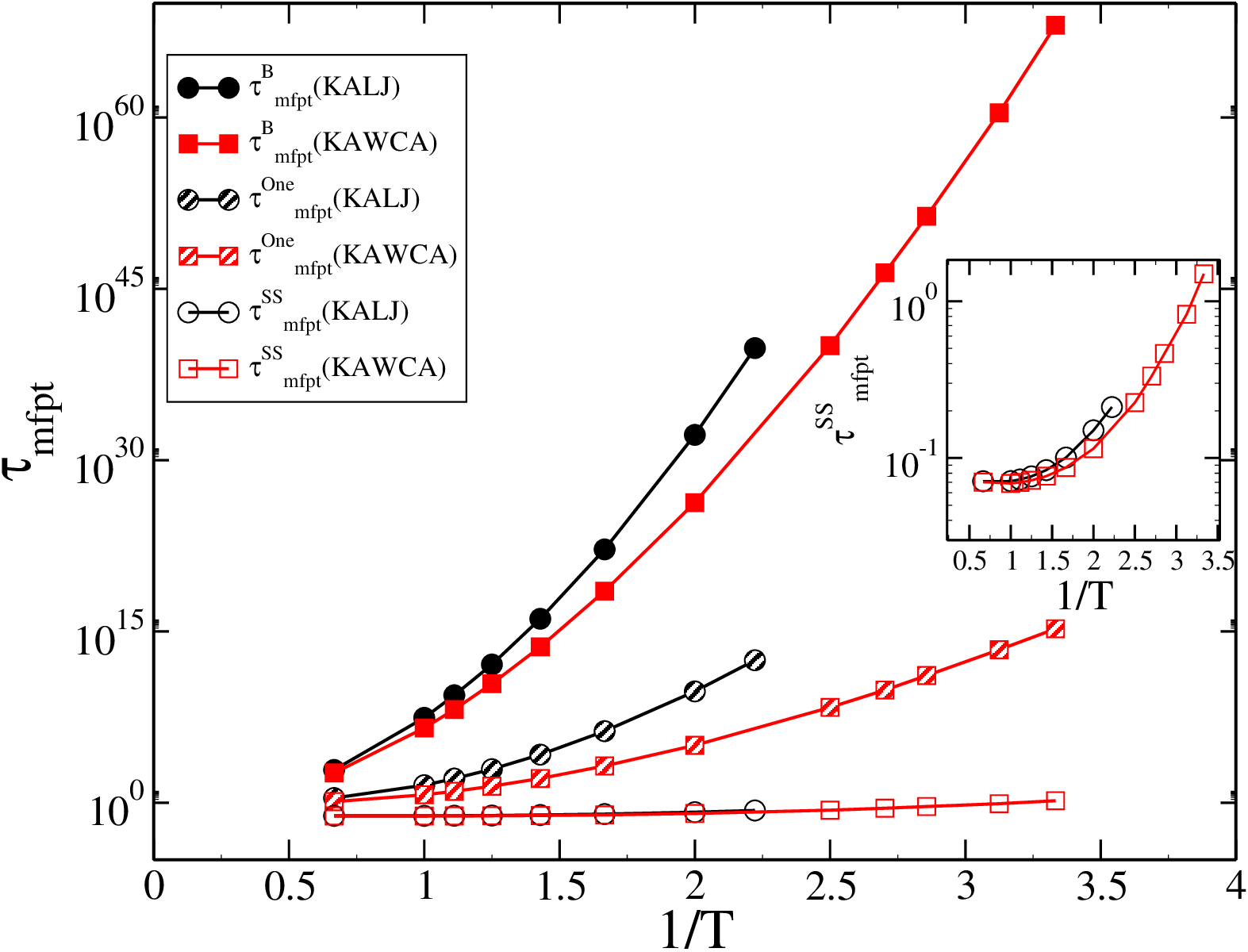}
	\caption{\it{ The Arrhenius plot of ${\tau_{mfpt}}$ against inverse temperature : The circles are for KALJ system and squares are for KAWCA system. The filled symbols are for binary systems (Eq.\ref{Eq.17}), the striped symbols are for one component description of binary systems (Eq.\ref{Eq.26}), the open symbols are for Schweizer-Saltzman systems (Eq.\ref{Eq.25}) and the inset shows only the $\tau^{SS}_{mfpt}$ values. The lines are guide to the eye.}}
	\label{fig.4}
\end{figure}

\begin{figure}[ht!]
	\centering
	\subfigure{
		\includegraphics[width=0.45\textwidth]{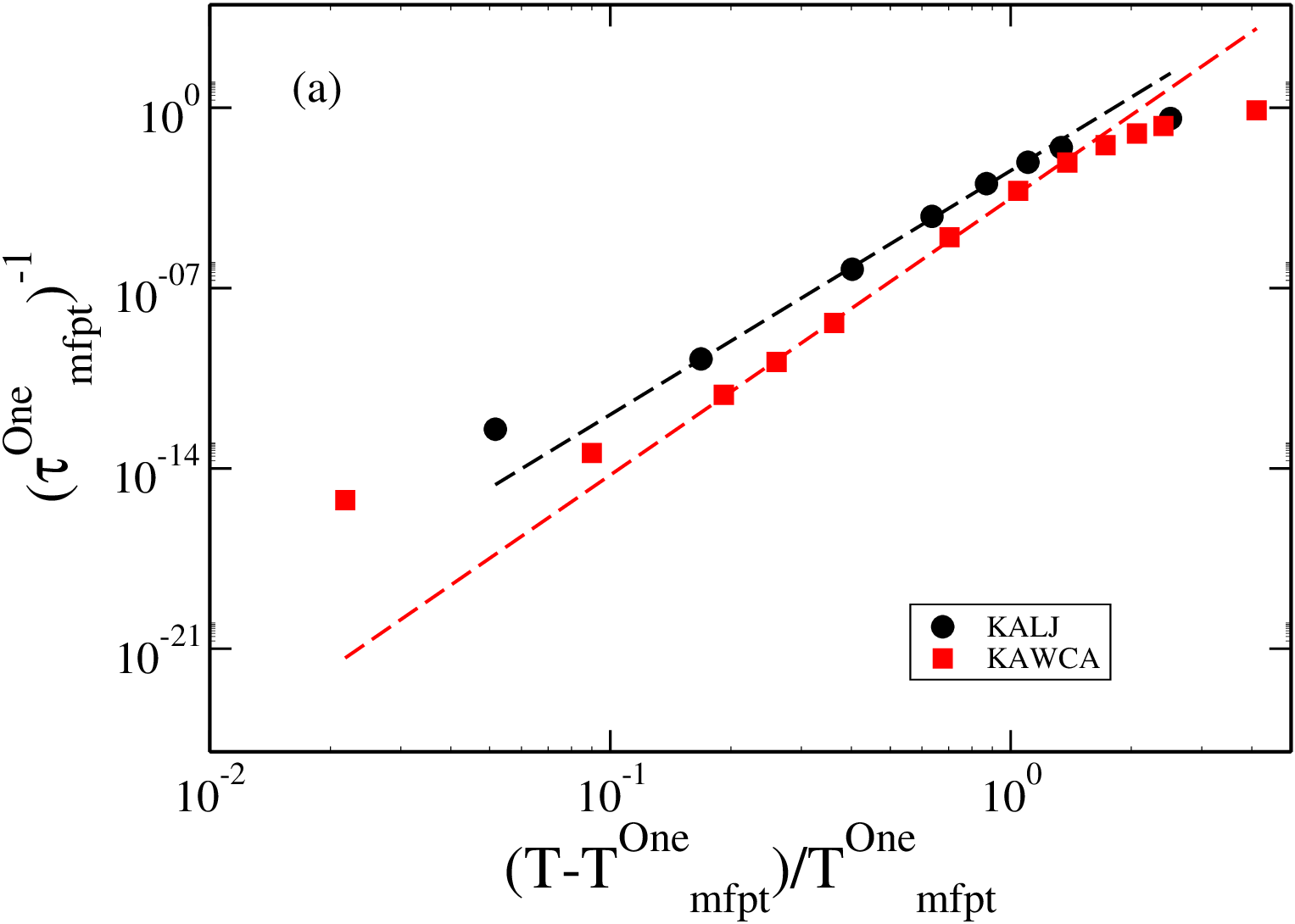}}
	\subfigure{
		\includegraphics[width=0.45\textwidth]{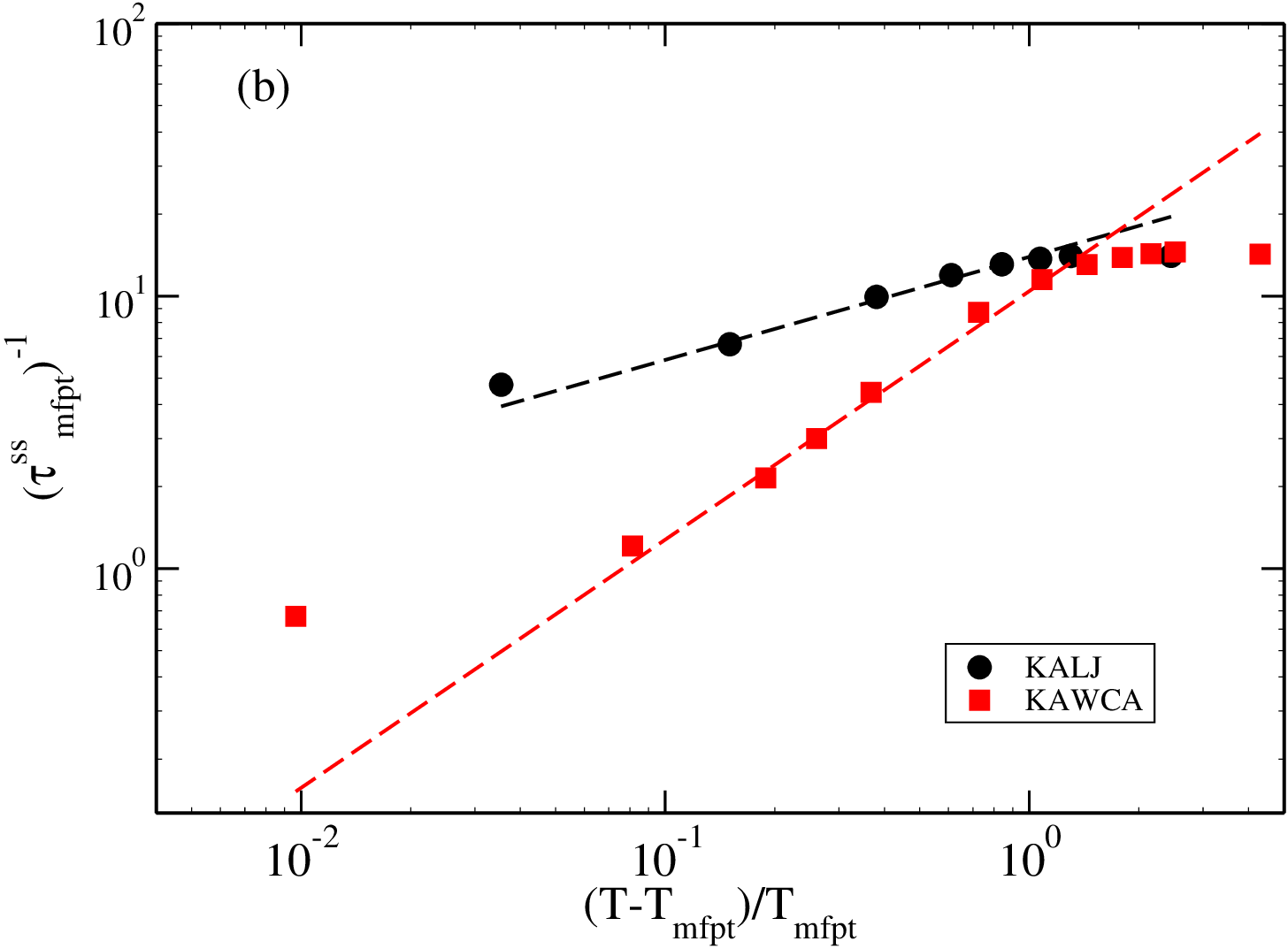}}
	\caption{\it{ The MCT power law fitting for the KALJ and KAWCA systems (a) For the mfpt relaxation time obtained in the one component version of the binary system, ${T^{One}_{mfpt}}$= 0.4279 (LJ) and 0.2936 (WCA), Here, the dashed lines are MCT fits. (b) For the mfpt relaxation time obtained from the Schweizer-Saltzman formalism, ${T^{SS}_{mfpt}}$= 0.4346 (LJ) and 0.2973 (WCA), Here, the dashed lines are MCT fits.}}.
	\label{fig.5}
\end{figure}

Next, we analyze why the Schweizer-Saltzman formalism \cite{Schweizer} failed to find the difference between the dynamics in the KALJ and KAWCA systems\cite{berthier_EPJE}. We have already shown that qualitatively $\tau_{Kramers}$ and $\tau_{mfpt}$ show similar results. Thus in the rest of the article, we concentrate on $\tau_{mfpt}$. Also, note that between our present formalism and Schweizer-Saltzman formalism there are a few differences. We work with binary system whereas Schweizer-Saltzman formalism is based on a one-component system and also there is a difference between the one component version of our theory and that used by Schweizer-Saltzman formalism. Thus to pinpoint which of these two factors allow our theory to predict the difference between the KALJ and the KAWCA systems, in Fig.\ref{fig.4} we plot the $\tau_{mfpt}$ values of the two systems as obtained for the binary system, $\tau^B_{mfpt}$, the one component version of the binary system (Eq.\ref{Eq.26}), $\tau^{one}_{mfpt}$ and also that predicted by the Schweizer-Saltzman formalism \cite{Schweizer}, $\tau^{SS}_{mfpt}$. Here although we use Schweizer-Saltzman formalism, instead of $\tau^{SS}_{Kramers}$ which was used in the original theory, we work with $\tau^{SS}_{mfpt}$. Similar to that observed for $\tau^{SS}_{Kramers}$ we find that $\tau^{SS}_{mfpt}$ for the LJ and the WCA systems are quite close. This finding corroborates the earlier observation \cite{berthier_EPJE} and strengthens the idea that qualitatively both the first passage times provide similar results. Interestingly we find that the difference in the dynamics between the KALJ and KAWCA systems is present in the one component version of our binary system. As mentioned before, in the Schweizer-Saltzman formalism, the free energy functional in Eq.{\ref{Eq.22}} was derived in a heuristic manner to ensure that the force obtained from the free energy matches the IMCT expression of the force. This gave rise to the $[1+S^{-1}(q)]$ term in the vertex and in the exponential. Also we have discussed that no such term can be present in the vertex in our free energy expression even if we avoid the Vineyard approximation. We analyzed the role of $[1+S^{-1}(q)]$ term in the exponential and in the vertex of the free energy. Our analysis reveals that the presence of $[1+S^{-1}(q)]$ term in the vertex, which does not have a microscopic origin actually blurs the difference between the LJ and WCA systems. 
\\ \\
We next check if the prediction of the difference in dynamics is good enough in predicting the difference in the MCT transition temperatures. Again in Tables \ref{t7}-\ref{t10}, we present the results as obtained using different fitting zones and then we use the best fitting and plot it in Fig.\ref{fig.5}. We find that similar to the binary dynamics the one component version of our present theory can also predict transition at a temperature which is closer to the MCT transition temperature. Thus the transition temperatures predicted for the KALJ and KAWCA systems by the one component version are also quite apart. However, the $\gamma$ values in the one component version are smaller than those for the binary system as due to the omission of certain terms the free energy minimum is shallower and the temperature dependence of the dynamics is weaker. The most surprising result is that even for the dynamics obtained using the Schweizer-Saltzman formalism the predicted transition temperatures for the two systems are quite apart and the values are close to the MCT transition temperatures obtained from the simulation studies (Tables \ref{t9}-\ref{t10} and Fig.\ref{fig.5}). This shows that although the absolute value of the relaxation time appears to be close between KALJ and KAWCA systems; the temperature dependence of them are quite different, which leads to the prediction of the difference in the transition temperatures.\\

Since the dynamics of the two systems are quite close, the difference in the transition temperatures might appear due to the difference in the fitting ranges of the two systems.
However, in the last line of Table \ref{t10} we report the parameters obtained for the KAWCA system when fitted in the temperature range 0.5-0.9 which is the best fitting zone for the KALJ system. The parameters clearly show that the transition temperature is close to the $T_{c}$ value of the KAWCA system. However, since this is not the best fitting zone for the WCA system, the MCT power law fit (not shown here ) is not good.
\\

The fact that all the different formalism can predict transition temperatures which are close to the $T_{c}$ value is quite remarkable. Note that in our fitting procedure there is no bias towards any particular value of $T_{c}$. However, we do not put any restriction on the $\gamma$ value as we do not expect any formalism which considers only up to the pair correlation to provide the actual dynamics and thus the correct temperature dependence of the system. Another observation from the present study is that all these formalism not only predict similar MCT transition temperature but the MCT fitting zone $(10^{-1}-10^{0})$ also remains unaltered.

\begin{table}
	\begin{center}
		\caption{\it{The fitted values of transition temperatures, $T^B_{mfpt}$, power law exponent, $\gamma$ and quality of the fit, $\chi^2$ (see text) for $\tau^B_{mfpt}$ (Eq.\ref{Eq.17}) in binary KALJ system. The fitting is performed over different temperature regimes for $\rho$ =1.2.  Here the smallest value of $\chi^2$ describes the best fitting.}} 
		
		\begin{tabular}{ |p{2.5cm}||p{2.5cm}|p{3cm}|p{2cm}| }
			\hline
			Region & $\chi^2$ & $T_{mfpt}^B{(LJ)}$ & $\gamma{(LJ)}$ \\
			\hline
			0.45-0.9 & $7.92\times10^{-6}$ & 0.3786 & 35.9101 \\
			0.45-0.8 & $3.59\times10^{-4}$  & 0.3671 & 39.0470 \\
			0.50-0.9 & $5.83\times10^{-7}$  & 0.4139 & 30.6156 \\
			0.50-0.8 & $1.13\times10^{-5}$  & 0.4010 & 33.3339 \\
			\hline
		\end{tabular}
		\label{t3}
	\end{center}
\end{table}

\begin{table}
	\begin{center}
		\caption{\it{The fitted values of transition temperatures, $T^B_{mfpt}$, power law exponent, $\gamma$ and quality of the fit, $\chi^2$ (see text) for $\tau^B_{mfpt}$ (Eq.\ref{Eq.17}) in binary KAWCA system. The fitting is performed over different temperature regimes for $\rho$ =1.2.  Here the smallest value of $\chi^2$ describes the best fitting.}} 
		
		\begin{tabular}{ |p{2cm}||p{2cm}|p{3cm}|p{2cm}|}
			\hline
			Region & $\chi^2$ & $T_{mfpt}^B{(WCA)}$ & $\gamma{(WCA)}$ \\
			\hline
			0.32-0.7 & $2.29\times10^{-6}$ & 0.2637 & 53.8966 \\
			0.32-0.6 & $3.31\times10^{-3}$ & 0.2544 & 58.9031 \\
			0.35-0.7 & $7.29\times10^{-8}$ & 0.2870 & 47.0907 \\
			0.35-0.6 & $2.61\times10^{-5}$ & 0.2755 & 51.9233 \\
			\hline
		\end{tabular}
		\label{t4}
	\end{center}
\end{table}

\begin{table}
	\begin{center}
		\caption{\it{ The fitted values of transition temperatures, $T_{Kramers}^B$, power law exponent, $\gamma$ and quality of the fit, $\chi^2$ (see text) for $\tau^B_{Kramers}$ (Eq.\ref{Eq.18}) in binary KALJ system. The fitting is performed over different temperature regimes for $\rho$ =1.2. Here the smallest value of $\chi^2$ describes the best fitting.}}
		
		\begin{tabular}{ |p{2cm}||p{2.5cm}|p{3cm}|p{2cm}| }
			\hline
			Region & $\chi^2$ & $T_{Kramers}^B{(LJ)}$ & $\gamma{(LJ)}$ \\
			\hline
			0.45-0.9 & $4.55\times10^{-7}$ & 0.3772 & 36.9321 \\
			0.45-0.8 & $2.03\times10^{-5}$  & 0.3658 & 40.0929 \\
			0.50-0.9 & $3.43\times10^{-8}$  & 0.4119 & 31.6385 \\
			0.50-0.8 & $6.69\times10^{-7}$  & 0.3877 & 34.3915 \\
			\hline
		\end{tabular}
		\label{t5}
	\end{center}
\end{table}

\begin{table}
	\begin{center}
		\caption{\it{The fitted values of transition temperatures, $T^B_{Kramers}$, power law exponent, $\gamma$ and quality of the fit, $\chi^2$ (see text) for $\tau^B_{Kramers}$ (Eq.\ref{Eq.18}) in binary KAWCA system. The fitting is performed over different temperature regimes for $\rho$ =1.2.  Here the smallest value of $\chi^2$ describes the best fitting.}} 
		
		\begin{tabular}{ |p{2cm}||p{2cm}|p{3cm}|p{2cm}|}
			\hline
			Region & $\chi^2$ & $T_{Kramers}^B{(WCA)}$ & $\gamma{(WCA)}$ \\
			\hline
			0.32-0.7 & $1.21\times10^{-7}$ & 0.2630 & 54.9266 \\
			0.32-0.6 & $1.71\times10^{-4}$ & 0.2537 & 56.9624 \\
			0.35-0.7 & $3.81\times10^{-9}$ & 0.2862 & 48.0662 \\
			0.35-0.6 & $1.32\times10^{-6}$ & 0.2748 & 52.9126 \\
			\hline
		\end{tabular}
		\label{t6}
	\end{center}
\end{table}
\noindent

\begin{table}
	\begin{center}
		\caption{\it{The fitted values of transition temperatures, $T^{One}_{mfpt}$, power law exponent, $\gamma$ and quality of the fit, $\chi^2$ (see text) for $\tau^{One}_{mfpt}$ (Eq.\ref{Eq.26}) in one component reduction of binary KALJ system. The fitting is performed over different temperature regimes for $\rho$ =1.2. Here the smallest value of $\chi^2$ describes the best fitting.}}
		
		\begin{tabular}{ |p{2cm}||p{2cm}|p{3cm}|p{2cm}| }
			\hline
			Region & $\chi^2$ & $T_{mfpt}^{One}{(LJ)}$ & $\gamma{(LJ)}$ \\
			\hline
			0.45-0.9 & $1.41\times10^{-1}$ & 0.3896 & 11.4587 \\
			0.45-0.8 & $9.37\times10^{-1}$ & 0.3783 & 12.5768 \\
			0.50-0.9 & $2.84\times10^{-2}$ & 0.4279 & 9.4681 \\
			0.50-0.8 & $1.64\times10^{-1}$ & 0.4142 & 10.4452 \\
			\hline
		\end{tabular}
		\label{t7}
	\end{center}
\end{table}

\begin{table}
	\begin{center}
		\caption{\it{The fitted values of transition temperatures, $T^{One}_{mfpt}$, power law exponent, $\gamma$ and quality of the fit, $\chi^2$ (see text) for $\tau^{One}_{mfpt}$ (Eq.\ref{Eq.26}) in one component reduction of binary KAWCA system. The fitting is performed over different temperature regimes for $\rho$ =1.2. Here the smallest value of $\chi^2$ describes the best fitting.}}
		
		\begin{tabular}{ |p{2cm}||p{2cm}|p{3cm}|p{2cm}|}
			\hline
			Region & $\chi^2$ & $T_{mfpt}^{One}{(WCA)}$ & $\gamma{(WCA)}$ \\
			\hline
			0.32-0.7 & $4.91\times10^{-1}$ & 0.2696 & 12.4461 \\
			0.32-0.6 & $0.45\times10^{1}$ & 0.2605 & 13.6717 \\
			0.35-0.7 & $1.42\times10^{-1}$ & 0.2936 & 10.7429 \\
			0.35-0.6 & $0.11\times10^{1}$  & 0.2822 & 11.9304 \\
			\hline
		\end{tabular}
		\label{t8}
	\end{center} 
\end{table}

\begin{table}
	\begin{center}
		\caption{\it{The fitted values of transition temperatures, $T^{SS}_{mfpt}$, power law exponent, $\gamma$ and quality of the fit, $\chi^2$ (see text) for $\tau^{SS}_{mfpt}$ (Eq.\ref{Eq.25}) in Schweizer-Saltzman KALJ system. The fitting is performed over different temperature regimes for $\rho$ =1.2.  Here the smallest value of $\chi^2$ describes the best fitting.}}

		\begin{tabular}{ |p{2cm}||p{2cm}|p{3cm}|p{2cm}| }
			\hline
			Region & $\chi^2$ & $T_{mfpt}^{SS}{(LJ)}$ & $\gamma{(LJ)}$ \\
			\hline
			0.45-0.9 & $2.0\times10^{-1}$ & 0.4201 & 0.4015 \\
			0.45-0.8 & $4.8\times10^{-1}$ & 0.4108 & 0.4576 \\
			0.50-0.9 & $1.8\times10^{-1}$ & 0.4346 & 0.3781 \\
			0.50-0.8 & $9.6\times10^{-1}$ & 0.3868 & 0.5562 \\
			\hline
		\end{tabular}
		\label{t9}
	\end{center}
\end{table}

\begin{table}
	\begin{center}
		
		\caption{\it{The fitted values of transition temperatures, $T^{SS}_{mfpt}$, power law exponent, $\gamma$ and quality of the fit, $\chi^2$ (see text) for $\tau^{SS}_{mfpt}$ (Eq.\ref{Eq.25}) in Schweizer-Saltzman KAWCA system. The fitting is performed over different temperature regimes for $\rho$ =1.2.  Here the smallest value of $\chi^2$ describes the best fitting.}} 
		
		\begin{tabular}{ |p{2cm}||p{2cm}|p{3cm}|p{2cm}|}
			\hline
			Region & $\chi^2$ & $T_{mfpt}^{SS}{(WCA)}$ & $\gamma{(WCA)}$ \\
			\hline
			0.32-0.7 & $3.21\times10^{-1}$  & 0.2937 & 0.9173 \\
			0.32-0.6 & $8.97\times10^{-1}$  & 0.2859 & 1.0491 \\
			0.35-0.7 & $3.15\times10^{-1}$  & 0.2973 & 0.9120 \\
			0.35-0.6 & $0.10\times10^{1}$  & 0.2979 & 0.9805 \\
			\hline
		\end{tabular}
		\label{t10}
	\end{center}
\end{table}    
\noindent
\clearpage

\section{Conclusions}

The present study is motivated by and also a followup of a recently presented study by us \cite{Manojprl}. We had shown that the dynamics of a system, when described at the mean-field level by retaining terms only up to pair correlations, can predict the dynamical transition temperature. Further, this microscopic mean field formalism could differentiate between the dynamics of the  KALJ and KAWCA systems. This leads us to conclude that the information of the MCT transition temperature is embedded in the pair structure of the liquid.
Interestingly our work is quite similar to the DFT formalism developed earlier by Schweizer and Saltzman\cite{Schweizer}. But the predictions of the two theories were quite different, as the Schweizer-Saltzman formalism failed to find the difference in the dynamics of the two systems \cite{berthier_EPJE}. There are certain differences present in the two theoretical formalism. Our earlier work described the dynamics on a potential energy surface whereas Schweizer-Saltzman formalism worked on the free energy surface. Our study was for a binary system where we started from the Fokker-Planck equation which was then reduced to a one-dimensional Smoluchowski equation and then the dynamics were obtained via mean first passage time. In the Schweizer-Saltzman formalism, the system was one component, and they used over-damped Langevin equation to describe the dynamics in an effective free energy surface. The timescale was obtained from the Kramers theory. 
Also in our formalism, we have made Vineyard approximation whereas in the Schweizer-Saltzman theory this approximation was not made. Note that a Fokker-Planck equation and a Langevin equation can be recast onto each other. Also as shown by Zwanzig in certain limit the Kramers barrier crossing time can be derived from the mean first passage time \cite{zwanzig2001nonequilibrium}. So it is difficult to understand why the prediction by the two formalism are different, and this is what is investigated in this work. \\

To close the gap between our work and Schweizer-Saltzman formalism, in this article we present the dynamics on the free energy surface of a binary system. The approximations used to describe the free energy surface are quite similar to those used in the earlier work to describe the potential energy surface \cite{Manojprl}. In the theoretical section, we present the derivation of the different first passage times. For the binary system, we derive the mean first passage time dynamics in the free energy surface and from there derive the Kramers first passage time. We then recapitulate the Schweizer-Saltzman formalism and show that we can recast their Langevin dynamics onto a one dimensional Smoluchowski equation and describe mean first passage time which again under certain approximations can be converted into Kramers first passage time. We also discuss the approximations required to bridge our free energy surface and Schweizer-Saltzman free energy surface. If we make a Vineyard approximation, on the Schweizer-Saltzman formalism, its expression becomes the one component version of our binary system. \\

We first show that both Kramers first passage dynamics and mean first passage time dynamics are qualitatively similar. They both can differentiate between the dynamics of the KALJ and the KAWCA systems and also can predict transition temperatures which are quite close to the MCT transition temperatures of the respective systems. In the rest of the article, we deal only with mfpt dynamics. We show that in the Schweizer-Saltzman formalism the mfpt dynamics of the KALJ and the KAWCA systems are quite close which is similar to that obtained for the Kramers first passage dynamics \cite{berthier_EPJE}. However, we find that the mfpt dynamics obtained from the one component version of our formalism can predict the difference between the two systems. Thus, we conclude that the $[1+S^{-1}(q)]$ term in the vertex in Schweizer-Saltzman theory is responsible for blurring the difference in dynamics between the two systems. We then show that even the one component version of our binary system can predict the MCT transition temperatures of both the systems. 

One of the interesting and unexpected results is that the dynamics for the two systems obtained from Schweizer-Saltzman formalism appear quite similar but when fitted to a MCT power law they predict the transition temperatures which are close to the respective $T_{c}$ values. 

In the present study, we also find that the MCT validity regime in all the different formalism does not change. Note that our fitting does not bias towards any known value of transition temperature. Thus the fact that in every case wherever we use the information of the pair structure (even the approximate one component version of it) we recover the transition temperature and the power law fitting regime does imply that the pair structure has the information of the MCT transition temperature. However, the magnitude of the relaxation time and its temperature dependence are not reproduced in our approximate calculations and this is reflected in the $\gamma$ values. These results are similar to that obtained in a different study by some of us where we have calculated the pair dynamics via the Adam-Gibbs expression using only the information of the pair configurational entropy\cite{Atryeeprl}. There we found that the dynamics vanishes at a temperature which is similar to the MCT transition temperature, but the temperature dependence of the dynamics was much stronger. For the systems discussed here the high temperature dynamics is primarily described by the pair correlation.  However at low temperature it is well known that many body correlation in the form of activation contributes to the total dynamics of the system. Thus any theory like the present one which ignores these many body correlations cannot describe the correct dynamics at low temperature. What it is able to predict is that $T_{c}$ appears to be the temperature where the high temperature dynamics (in these systems described by the pair correlation) disappears.

\begin{thebibliography}{10}
\bibitem{correction}
	{ In our earlier work \cite{Manojprl} there was a mistake in the expression
		of the potential energy term given in Eq.2 of the main manuscript and Eq.10
		in the supplimentary{.} The correct expression for a one component system
		is{,} $ \phi(r) \equiv -\int \frac{dq}{(2\pi)^3}\rho
		C^2{(q)}S(q)exp-[\frac{q^2r^2}{6}]$. However, the conclusion of the paper
		remains unchanged as we find that the correct expression can also predict the
		MCT transtion temperature.} .
\end{thebibliography}

\end{document}